# Strip-Loaded Nanophotonic Interfaces for Resonant Coupling and Single-Photon Routing

Katharine Snow, Fatemeh Moradiani, and Hamidreza Siampour*

School of Mathematics and Physics, Queen's University Belfast, University Road, Belfast, BT7 1NN, UK

* h.siampour@qub.ac.uk

**Abstract**: We report on the design and simulation of strip-loaded nanophotonic interfaces aimed at improving resonant coupling and photon routing efficiency. In our design, the guided mode is confined within a plane by a high-index thin film and is loosely confined laterally by a lower index strip. Using a hydrogen silsesquioxane (HSQ) strip, titanium dioxide core, and silicon dioxide substrate, we optimise the waveguide dimensions for maximum lateral confinement of light. Specifically, we propose a polymer-based Bragg grating cavity and ring resonator that achieve near-optimal mode volumes and high Q-factors. Our simulations suggest that a cavity with a mode volume of $V_{eff} \sim 7.0(\lambda/n)^3$ and a Q-factor of 7000 can produce photons with 97% indistinguishability at 4K. Additionally, we investigate directional couplers for efficient photon routing, comparing photonic and plasmonic material structures. While pure photonic structures demonstrate lower loss and improved quality factors, they face practical limitations in terms of bending radius. Conversely, plasmonic structures offer shorter bending radii but higher propagation losses. This research lays the groundwork for future nanophotonic designs, aiming to enhance photon generation and routing capabilities for quantum optical applications.

**Introduction**

Quantum emitters of single indistinguishable photons are critical for advancing quantum technologies such as computing (*1*), networking (*2*), cryptography (*3*), and metrology (*4*). Currently, spontaneous parametric down-conversion is the prevalent method for single-photon generation, but its probabilistic nature limits both the efficiency and indistinguishability of photons (*5*). Solid-state emitters, including semiconductor quantum dots (*6, 7*) and diamond colour centres (*8, 9*), offer promising alternatives by enhancing these properties when integrated with nanophotonic devices. For example, by placing the emitters inside a high-Q cavity, the emitter's decay rate is enhanced by the Purcell factor $F_P$, which reduces the effects of dephasing and decay to unwanted modes. Unlike placing the emitter in a waveguide, using a cavity also filters out the non-Markovian phonon sideband without sacrificing efficiency (*10*). This is crucial when using emitters with a large sideband, as achieving near-unity efficiency is a primary research focus due to the stringent requirements imposed by all potential applications.

To achieve narrowband enhancement of the zero-phonon line (ZPL), the cavity decay rate $\kappa$ must satisfy the condition $\kappa \ll \xi$ where $\xi$ is the typical energy scale of coupled phonons. The single figure of merit is the Purcell factor, which is engineered to be as high as possible, subject to the constraint of weak coupling. The Purcell factor is given by:

$$F_P = \frac{2g^2}{\Gamma \kappa} = \frac{3}{4\pi^2} \frac{Q}{V_{eff}/(\lambda/n)^3}; \qquad (1)$$

where $g$ is the emitter-cavity coupling rate, $\Gamma$ is the emitter's vacuum decay rate, $\kappa$ is the cavity decay rate, $Q$ is the cavity quality factor, and $V_{eff}$ is its mode volume. Increasing the Purcell factor therefore means increasing the coupling $g$ while keeping $\kappa$ just within the limit of weak coupling,

$\kappa/2 \geq g$. In more practical terms, this equates to reducing the mode volume as far as possible and increasing $Q$ to just within the weak coupling limit. The coupling cannot be increased arbitrarily, however, without losing the narrowband enhancement, $\kappa \ll \xi$. There has been a focus on designing cavities with deeply sub-wavelength mode volumes, which are theoretically achievable using slot waveguides or plasmonic devices (*11-13*), and employing both direct and inverse design approaches (*14-16*). However, experimentally, the best-performing devices use quantum dots embedded in conventional Bragg cavities, with mode volumes on the order of $V_{eff} \sim (\lambda/2n)^3$, and lossy plasmonic components are avoided. Quantum dots in micropillar cavities at cryogenic temperatures have been used to generate single photons with indistinguishabilities greater than 0.99 and efficiencies around 0.9—excluding inefficiencies in out-coupling or initial excitation—which has enabled their use in generating small photonic cluster states of approximately 10 qubits (*17, 18*).

Micropillar cavities are chosen for their high $Q$-factors upon fabrication. The difficulty of precise nanoscale positioning of the emitter within the cavity also prohibits the use of ultra-small mode volumes. Under the constraints $\kappa/2 \geq g$ and $\kappa \ll \xi$, there is a maximum Purcell enhancement of $F_P \Gamma \sim \xi/20$ for the optimal values $\kappa \sim \xi/10$ and $g \sim \xi/20$. For quantum dots with $\xi \sim 1.45$ me, we demonstrate that this corresponds to $Q \sim 7000$, $V_{eff} \sim 7.0(\lambda/n)^3$, and $F_P \sim 50$.

Here, we propose a polymer-based Bragg grating cavity and ring resonator. Such polymer-based nanophotonic devices can be fabricated with $Q$-factors greater than $10^4$ using current fabrication technology (*9, 19*). Due to the low refractive index of hydrogen silsesquioxane (HSQ) polymer, we choose a strip-loaded waveguide geometry, which results in a larger mode volume than current devices, with $V_{eff} \sim 7.0(\lambda/n)^3$. When interfaced with a quantum dot at 4K, we predict that a cavity with this mode volume, if fabricated with $Q \sim 7000$, will generate photons with an indistinguishability of 97% and an efficiency of 98%. Our cavity provides a model on which to base further design improvements to achieve the necessary very high $Q$-factors and low footprint in these devices, within the planar configuration that makes them suitable for on-chip manipulation of the emitted photons.

**Results and Discussion**

**Bragg grating cavity**
In a strip-loaded waveguide, the mode is confined within a plane by a high-index thin film and is loosely confined in the lateral direction by a lower index strip, as shown in Figure 1a(i). We choose an HSQ strip, titanium dioxide core, and silicon dioxide substrate, and optimise the waveguide dimensions for maximum lateral confinement of light with a vacuum wavelength of approximately 600 nm. The optimal dimensions are as follows: strip height of 180 nm, strip width of 300 nm, and core thickness of 70 nm. A Bragg cavity is constructed based on this waveguide geometry, with the Bragg reflectors being twice as long as the waveguide is wide, as shown in the inset of Figure 1a(ii). The transmission spectrum of this cavity is shown in Figure 1a(ii) for different values of the total number of Bragg reflectors *N*. For *N*=100, the quality factor is found to be 230. The cavity mode is shown in Figure 1c, and the mode volume is found to be 6.96 $(\lambda/n)^3$. We also show the stopband in Figure 1a(iii) and, for comparison, in Figure 1b the equivalent results for a plasmonic cavity with the same HSQ strip (*20*).

In the narrowband-enhancement weak-coupling regime (see Figure 2), the indistinguishability and efficiency of emitted photons are given by $I = \frac{F_P+1}{F_P+1+\frac{2\gamma}{\Gamma}}$, and $\eta = \frac{B^2 F_P}{B^2 F_P+1}$, respectively, where γ is the dephasing rate and $B$, known as the Frank-Condon factor, is the square-root probability of photon emission into ZPL (*10*). For example, for a quantum dot at 4K the optimal values of $Q$ and $V_{\text{eff}}$— corresponding to $\kappa \sim \xi/10$ and $g \sim \xi/20$—are given by $Q = \frac{\omega}{2\kappa}$, and $V_{\text{eff}} = \frac{\omega \mu^2}{2\hbar \epsilon_0 n^2 g^2}$, where $\omega$ is the angular frequency and $\mu$ the dipole moment. Taking $\mu = 27.2D$ (*21*), $2\pi c\omega$=600 nm and $\xi$ = 1.45 meV, we find the optimal values $Q$=7100 and $V_{\text{eff}}$ =7 $(\lambda/n)^3$. We should note that the wavelength used in these calculations is that of our proposed cavity, which differs from that of typical QDs by a factor of 1.5.

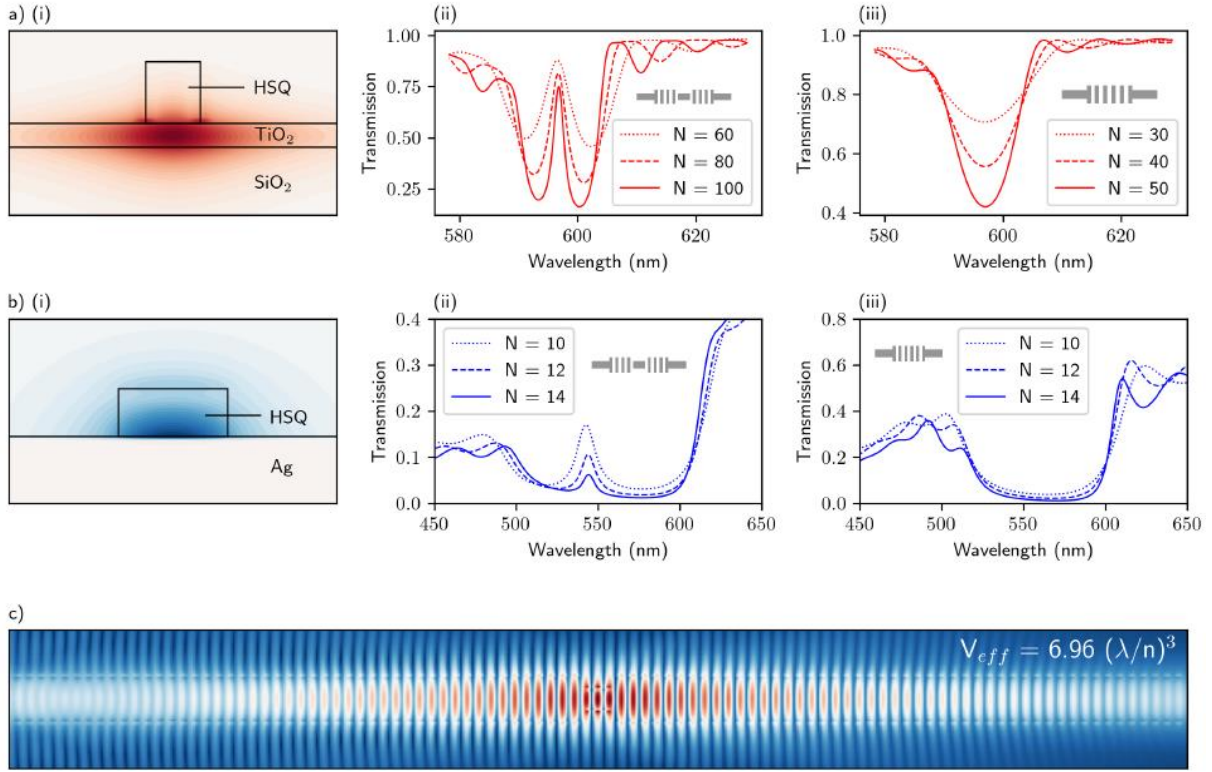

**Figure 1**: Electromagnetic simulations of polymer-based nanophotonic devices. (a) A strip-loaded waveguide and (b) a plasmonic waveguide. Cross-sections of the modes are depicted in (i). Transmission spectra for the corresponding Bragg cavities and Bragg reflectors are presented in (ii) and (iii), respectively, for various values of N, the number of reflectors. The strip-loaded cavity mode for N=100 is shown in (c).

To determine the indistinguishability, we assume that the dominant contribution to the dephasing $\gamma$ is due to phonons. This assumption holds in electrically contacted structures, where charge and spin noise are minimized (*22*). We also assume a bulk phonon spectrum, achievable through clamping (*23*). The dephasing then depends on $g$ and $\kappa$ through:

$$\gamma = 2\pi \left(\frac{gB}{\kappa}\right)^2 J(2gB) \coth\left(\frac{\hbar gB}{k_B T}\right); \tag{2}$$

where $J(\nu) = \alpha \nu^3 \exp\left(-\left(\frac{\nu}{\xi}\right)^2\right)$, $\alpha = 0.03 ps^2$ is the exciton–phonon coupling strength, $T$ is the temperature, and we take $B$=0.95 (*10*). In Figure 2, we show the combined value $I\eta$ as a function of $V_{\text{eff}}$ and $Q$. The regions outside the narrowband-enhancement weak-coupling regime have been excluded. At the point where $F_P$ is at its maximum, the indistinguishability is $I = 0.97$ and the efficiency is $\eta = 0.98$.

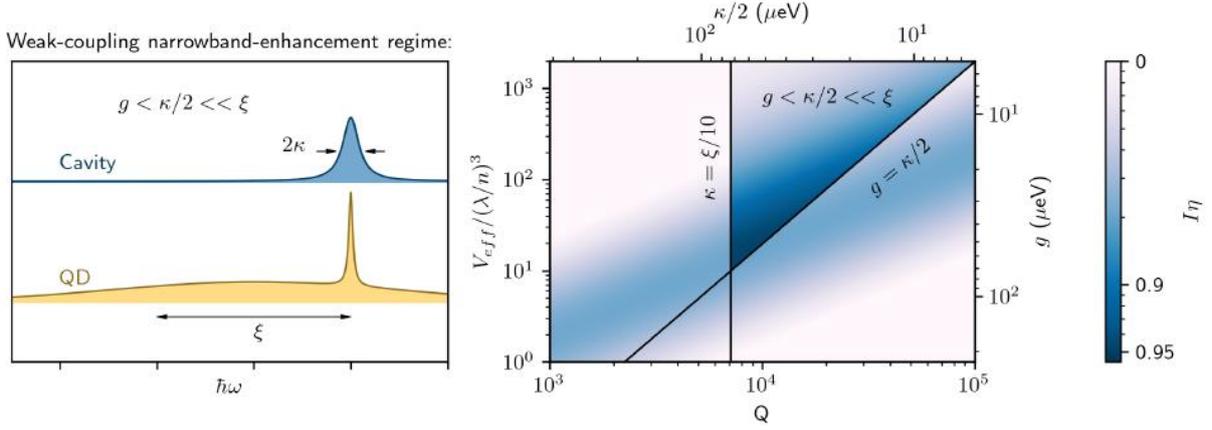

**Figure 2**: The narrowband-enhancement weak-coupling regime for a quantum dot embedded in a nanophotonic cavity. Left: A sketch illustrating the different energy scales of importance, where the cavity linewidth is much smaller than the typical energy scale of phonons. Right: The product $I\eta$ of the indistinguishability $I$ and the efficiency $\eta$ must be close to 1 for optimal quantum photonic applications. This product is plotted as a function of the $Q$-factor and mode volume. In our calculation, we assume that dephasing is primarily due to phonon effects and use typical experimental values for the coupling parameters in quantum dots.

**Directional coupler**

The concept of the directional coupler, introduced in the mid-20th century, has been foundational for many advancements in optical and microwave technologies (*24*). While the fundamental idea remains relevant, the design of the entire structure—including S-bends and a chain of materials tailored for specific applications—continues to hold significant importance in contemporary engineering. More recently, the directional coupler has become a crucial component in photonic quantum integrated circuits, demonstrating the capability to facilitate quantum computing logic operations when employed with single photons (*25-27*).

In this section, we present two platforms for directional couplers: one based on dielectric-loaded surface plasmon polariton (DLSPP) waveguides and the other on strip-loaded waveguide configurations. We then compare the two structures in terms of their applicability for single-photon routing. Notably, the overall dimensions of both platforms are consistent with the Bragg grating cavity structure discussed in the previous section.

The directional coupler is designed for both parallel straight waveguides and S-bends based on sine curves, enabling continuous bend curvature and adiabatic modification (*28-30*) of the DLSPP waveguide and the strip-loaded waveguide mode throughout the bend, as shown in Figure 3a(i) and 3b(i), respectively. One crucial aspect to consider is the design of the S-bend. As illustrated in Figure 3(a), the losses associated with the S-bend can be categorised into transition losses and radial losses.

Radial losses occur due to the limited guiding of the propagation mode caused by speed limitations beyond the outer end of the bent waveguide. Transition losses occur when there are discontinuities in the curvature of the waveguide, where sudden changes in modal propagation characteristics take place. The guided modes in curved waveguides are broader compared to straight waveguides and change outward along the curve. The normalised transition loss due to mismatch between the straight waveguide mode and the curvature can be calculated as follows (*29*):

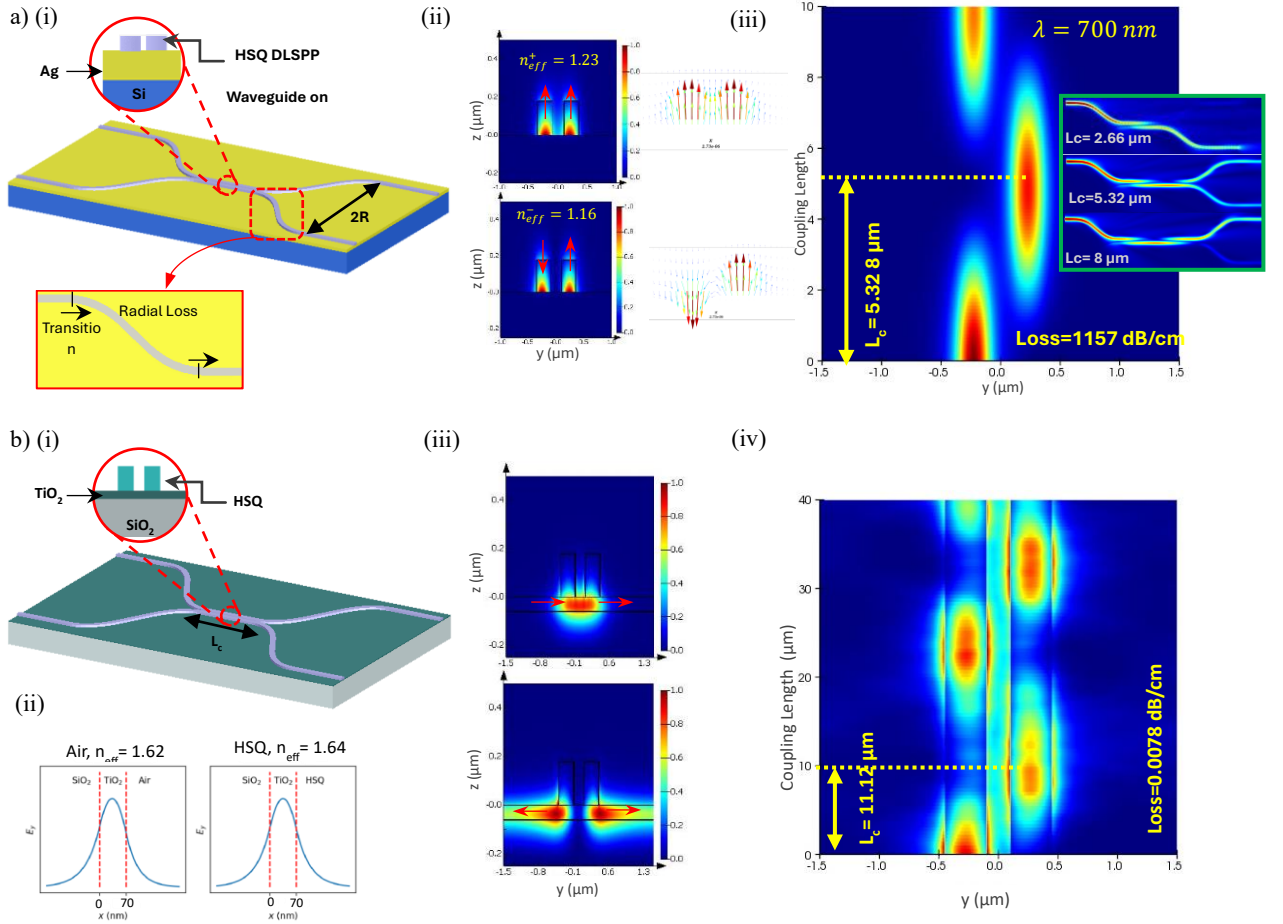

**Figure 3:** (a) Simulation results of a DLSPP-based directional coupler structure consisting of two rectangular DLSPP waveguides with a height of h=180 nm, width of w=250 nm, refractive index $n_d$=1.41, a separation gap of $g$=200 nm in the parallel section, and an S-bend waveguide with a 2.5 µm radius. The length of the parallel section (coupling length) is $Lc$=5.3 µm, designed to impart a π phase shift and achieve 50-50 power splitting at $\lambda$=700 nm between the (ii) symmetric ($n_{eff}^+$) and (iii) anti-symmetric ($n_{eff}^-$) DLSPP modes supported by (iv) the structure using $Lc=\lambda/2(n_{eff}^+ - n_{eff}^-)$. The surface shows the electric field norm (V/m) profile with red arrows indicating the E-field. Insets (ii): E-field mode vectors for symmetric and anti-symmetric modes. Inset (iv): The full profile of DLSPP directional couplers with a 200 nm gap and radii of 2.66 µm, 5.32 µm, and 8 µm, achieving 0-100, 50-50, and 100-0 power coupling, respectively. (b) Simulation results of the strip-loaded directional coupler structure consisting of two rectangular strip-loaded waveguides with a height of h=180 nm, width of w=250 nm, refractive index $n_d$=1.41, a separation gap of $g$=200 nm in the parallel section, and an S-bend waveguide with a 100 µm radius on a TiO₂ layer with a height of 70 nm. The length of the parallel section (coupling length) is $Lc$=11.2.

$$\Gamma_T = 1 - \frac{|\iint E_{in}(x,y) E_{out}^*(x,y) dy dx|^2}{\iint |E_{in}(x,y)|^2 \iint |E_{out}(x,y)|^2 dx dy}; \qquad (3)$$

where $E_{in}$ and $E_{out}$ represent the electrical input and output mode fields, respectively. The normalised radial loss, related to the radial attenuation coefficient $\alpha(R, l)$ per unit length $l$ in a bend radius $R$, can be expressed as (29): $\Gamma_R = exp\left(-\int_L^o \alpha(R(l'), l') dl'\right)$.

In addition to these losses, the effective refractive index resolution plays a critical role in the design of radial bends. Due to the high resolution in the DLSSP waveguide structure, the minimum radius required to guide light through the curve is approximately 2 µm. Higher field confinement and density of states can be achieved with SPPs, as they are interface modes and are not constrained by the diffraction limit. This enables smaller V$_{eff}$ and enhances both linear and nonlinear optical processes. However, the high energy density near the metal region inevitably leads to significant ohmic loss, which becomes more pronounced as the mode volume decreases (*31, 32*).

In the strip-loaded structure, the low refractive index of HSQ (n= 1.41) results in a minimal difference in the effective refractive index between the strip-loaded configuration with HSQ cladding and one without it, approximately 0.02, as shown in Figure 3b(ii). This difference leads to a minimum bending radius of around 100 µm. This characteristic is crucial for designing optical devices with tight bends and for sensitive optical applications, such as single-photon routing. However, in applications like single-photon sources, the overall propagation length is limited by structural and environmental factors when the emitter is coupled to the waveguide. Although the photon propagation length is generally longer than that of SPPs, issues may arise concerning the structure's overall length. Additionally, in these structures, distinguishability may be significantly compromised at very high dimensions. Nevertheless, the strip-loaded structure remains a strong option for applications where minimising loss is the primary concern.

The interaction length required to achieve the desired power coupling to the second waveguide is calculated using $L_c = \lambda/2(n_{eff}^+ - n_{eff}^-)(\sin^{-1}(\sqrt{P_{out2}/P_{in}}))$. For a π phase shift, the relation simplifies $L_c = \lambda/2(n_{eff}^+ - n_{eff}^-)$. Figure 3a(iii) and Figure 3b(iv) show the numerical calculations of the coupling length for both DLSPP and strip-loaded structures, using symmetric (Figure 3a(ii) top and Figure 3b(iii) top) and asymmetric modes (Figure 3a(ii) bottom and Figure 3b(iii) bottom) at a wavelength of 700 nm. While strip-loaded structures generally exhibit lower loss compared to DLSPP structures, their large radius of curvature can still limit their practical application in certain scenarios.

**Ring resonator**

Another approach to enhancing the Purcell effect is to use DLSPP waveguide-ring resonators (WRRs). A WRR, consisting of a straight waveguide laterally coupled to a ring resonator, typically provides more pronounced wavelength selection and a high *Q*-factor, which is crucial in integrated optics. It is important to note that the extinction ratio, which indicates the contrast in transmission through a WRR between resonant and non-resonant wavelengths, is affected by both the coupling efficiency and the internal loss within the resonator (*33*). In plasmonic WRRs, coupling efficiency is usually low due to the tight confinement in plasmonic waveguides, while internal loss is high due to propagation loss. As a result, achieving critical coupling for a high extinction ratio can be challenging.

Figure 4(a) shows a 3D schematic of the DLSSP-WRR structure, designed with the same parameters discussed in the previous section. Figure 4(b) presents the transmission spectrum for a ring with a radius of 2.5 µm and three different separations (g) of 200 nm, 150 nm, and 100 nm. A 100 nm gap provides a higher extinction ratio compared to the other gaps, though this enhanced coupling efficiency also leads to mode broadening. Figure 4(c) depicts the full profile of the structure at a resonant wavelength of 616.4 nm and a non-resonant wavelength of 623 nm. The bandwidth of the WRR is determined by the radius of the ring resonator and the wavelength-dependent effective index of the bent waveguide. Figure 4(d) displays the transmission spectra for a 100 nm separation and ring radii ranging from 2 µm to 3.5 µm in 0.5 µm increments. Increasing the ring radius improves the maximum resolution of the DLSSP, resulting in a higher quality factor. Figure 4(e) shows the mode area of the 250 nm width HSQ as a function of HSQ thickness (h). The mode area definition related to the Purcell effect, based on [10], indicates that as HSQ thickness increases from 160 nm to 240 nm, the effective mode area remains unchanged. However, for h<160 nm, mode leakage into the air cladding increases the mode area. Additionally, we have computed the mode area, normalized Purcell factor, and quality factor of the structure for various ring radii. As the ring radius increases, the bandwidth decreases, leading to an increase in the quality factor. For a ring with a radius of 2.5 µm, the Purcell factor and quality factor are ~10 and ~133, respectively, which are relatively high compared to a strip-loaded waveguide.

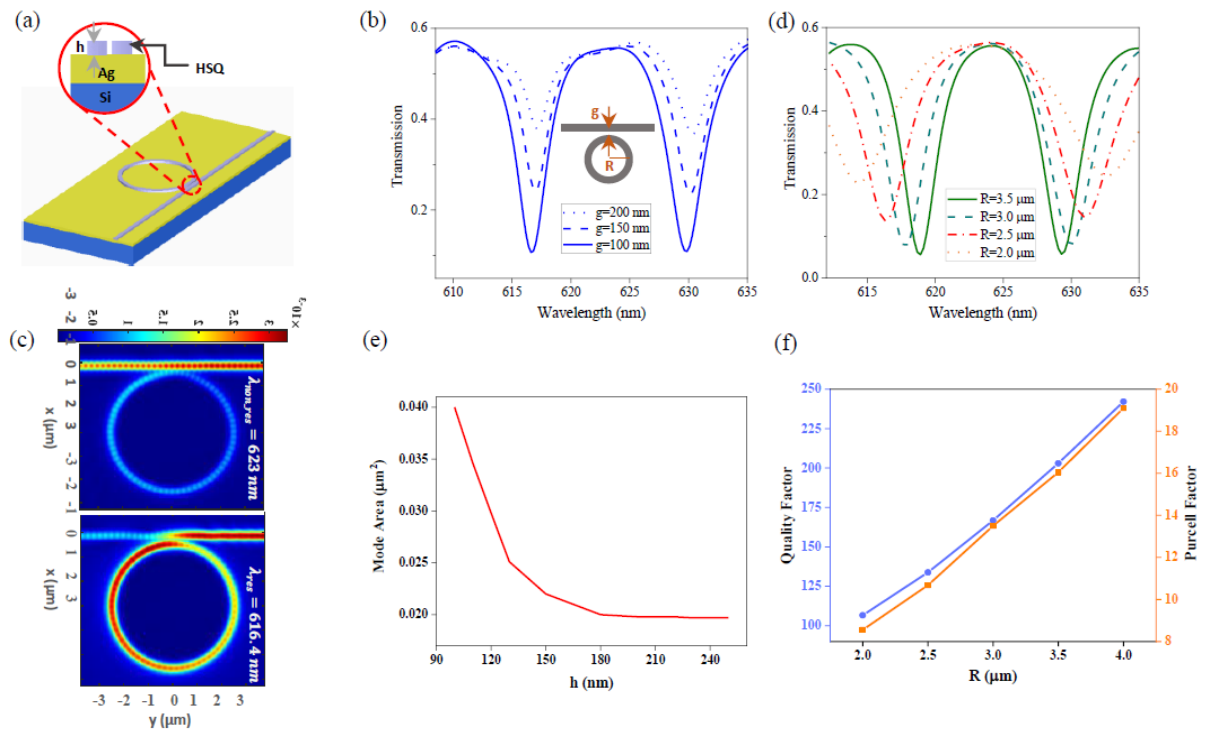

**Figure 4**: (a) Sketch of the DLSSP-WRR. (b) Transmission spectrum of the HSQ DLSPP-loaded ring-resonator cavity with a 2.5 µm radius, showing spectra for separation gaps of 200 nm, 150 nm, and 100 nm. (c) Full profile of the structure at the resonant wavelength of 616.4 nm and the non-resonant wavelength of 623 nm, for a separation gap of 100 nm. (d) Mode area as a function of HSQ thickness. (e) Calculated mode area as a function of HSQ thickness. (f) Quality factor and Purcell factor as functions of ring radius. In the transmission spectra (b), different ring radii (2.0 µm, 2.5 µm, 3.0 µm, and 3.5 µm) are displayed. The wavelength range from 610 nm to 635 nm is plotted to highlight the transmission characteristics at various separations and ring radii.

## Conclusion

We have introduced the design and simulation of a polymer-based Bragg grating cavity with a strip-loaded waveguide. The optimal parameters for this design include a mode volume of $V_{eff} \sim 7.0(\lambda/n)^3$, with predicted photon indistinguishability of 97% and efficiency of 98% for a cavity with a Q-factor of approximately 7000. Our analysis compared two types of directional couplers: DLSPP and strip-loaded waveguides, evaluating their design, performance, and limitations, with a focus on their applicability for single-photon routing. Additionally, we explored ring resonators for enhancing the Purcell effect, presenting simulations of a DLSSP-based ring resonator with various parameters. The results showed that increasing the ring radius improves resolution and quality factor but may lead to broader mode profiles. Our findings suggest that polymer-based cavities, specifically using HSQ, can achieve high Q-factors and significant Purcell factors. These structures are promising for practical quantum photonics applications due to their planar configuration, which facilitates on-chip integration. The proposed structures, including Bragg gratings and ring resonators, have demonstrated high potential for improving photon indistinguishability and efficiency, which are crucial for advancing quantum technologies.

**Acknowledgements**

This work was funded by the UKRI Strength in Places Fund programme Smart Nano NI, and the Engineering and Physical Sciences Research Council (EPSRC) under grant number EP/S023321/1.